\documentclass[pra,twocolumn]{revtex4-2}

\usepackage{amssymb}
\usepackage{graphicx}\usepackage{dcolumn}\usepackage{bm}\usepackage{physics}
\usepackage{float}
\usepackage{subfigure}
\usepackage{qcircuit}
\usepackage{hyperref}
\usepackage[usenames,dvipsnames]{xcolor}

\hypersetup{
    colorlinks,
    citecolor=blue,
    linkcolor=blue,
    urlcolor=blue
}

\begin{document}

\preprint{APS/123-QED}

\title{Avoiding Barren Plateaus with Entanglement}

\author{Yuhan Yao}
 \email{yao@biom.t.u-tokyo.ac.jp}
\author{Yoshihiko Hasegawa}\email{hasegawa@biom.t.u-tokyo.ac.jp}
\affiliation{Department of Information and Communication Engineering,
Graduate School of Information Science and Technology,
The University of Tokyo, Tokyo 113-8656, Japan
}\date{\today}

\begin{abstract}
In the search for quantum advantage with near-term quantum devices, navigating the optimization landscape is significantly hampered by the barren plateaus phenomenon. 
This study presents a strategy to overcome this obstacle without changing the quantum circuit architecture. 
We propose incorporating auxiliary control qubits to shift the circuit from a unitary $2$-design to a unitary $1$-design, mitigating the prevalence of barren plateaus. We then remove these auxiliary qubits to return to the original circuit structure while preserving the unitary $1$-design properties.
Our experiment suggests that the proposed structure effectively mitigates the barren plateaus phenomenon. 
A significant experimental finding is that the gradient of $\theta_{1,1}$, the first parameter in the quantum circuit, displays a broader distribution as the number of qubits and layers increases.
This suggests a higher probability of obtaining effective gradients.
This stability is critical for the efficient training of quantum circuits, especially for larger and more complex systems. The results of this study represent a significant advance in the optimization of quantum circuits and offer a promising avenue for the scalable and practical implementation of quantum computing technologies. This approach opens up new opportunities in quantum learning and other applications that require robust quantum computing power.
\end{abstract}

\maketitle

\section{Introduction\label{sec:introduction}}

Quantum information science, especially quantum computing, has made significant theoretical and experimental progress in recent years. With the advent of the first generation of quantum computers, we have entered the era of noisy intermediate-scale quantum (NISQ) devices \cite{endo2021nisq, bharti2022nisq, kjaergaard2020nisq}. Despite the potential advantages that these early quantum computers have in tackling complex computational problems, their practical application is hindered by several factors, including the quality of the qubits, error rates in the quantum gates, and algorithmic optimization problems such as the phenomenon of barren plateaus \cite{McClean2018, du2022bpvqa, patti2021bp, cerezo2022bp, shen2020bpqnn, wiersema2020bp, Russell2017bp}. The concept of ``barren plateaus" is important for machine quantum learning and optimization. It refers to regions in the parameter space of quantum neural networks \cite{shen2020bpqnn} whose gradients are minimal, close to zero. This poses a major challenge for gradient-based optimization methods, such as gradient descent, because the gradient information provides little to no guidance, leading to stagnation in the training process. This phenomenon is characterized by extremely flat gradients in high-dimensional quantum parameter spaces. It has proven to be a critical bottleneck in optimizing quantum algorithms and hinders the further development of quantum computing.

The implementation of quantum algorithms can be hindered by barren plateaus, which is especially problematic for quantum machine learning algorithms and variable quantum algorithms (VQAs) \cite{tilly2022vqa, cerezo2021vqa, du2022bpvqa, wecker2015vqa, Peruzzo2014vqa}. As quantum systems increase in size, the issue of gradient descent becomes more pronounced during the optimization process, resulting in inefficiencies in local search strategies. Moreover, this challenge is closely related to the limitations of quantum hardware, such as the limited number of qubits \cite{bergli2009limit_qubit, devoret2004limit_qubit, gambetta2017limit_qubit} and errors in quantum gate operations \cite{cai2021quantum_error, cai2023quantum_error, dur2007quantum_error, devitt2013quantum_error}. These challenges affect the practicality of quantum algorithms and limit the potential applications of quantum computing in various fields.

This study explores the causes and possible solutions of the barren plateaus phenomenon and its impact on optimizing quantum algorithms. We focus on developing new algorithms and strategies to address the issue of gradient vanishing in high-dimensional quantum parameter spaces.  This includes improving optimization strategies, exploring effective initialization protocols, and investigating the relationship between gradient vanishing and the properties of quantum hardware. Our ultimate goal is to enhance the efficiency and success rate of quantum algorithm optimization, thus enabling the application of quantum computing in practical scenarios. Additionally, we provide insights into the understanding and control of complex quantum systems.

In related works, we review the essential literature and theoretical foundations related to the phenomenon of barren plateaus in variational quantum algorithms (VQAs) \cite{du2022bpvqa, Sack2022, Marrero2021, Grant2019, Cerezo2021}. Barren plateaus are characterized by gradient vanishing issues in parameterized quantum circuits and have become a focal point in current quantum computing research. First, we refer to studies on the relationship between entanglement and learning efficiency in quantum neural networks \cite{Sack2022}, which suggest that excessive quantum entanglement can lead to reduced learning efficiency. Additionally, regarding the importance of initialization strategies in overcoming barren plateaus, we examine technical notes on new initialization techniques \cite{Marrero2021} that prevent early-stage gradient disappearance by controlling initial parameter selection. Next, we consider research defining weak barren plateaus (WBPs) based on the entropies of reduced local density matrices \cite{Grant2019}, providing a new perspective for understanding and quantifying barren plateaus. Lastly, we synthesize studies on the impact of observable selection on trainability in VQAs \cite{Cerezo2021}, highlighting the differences in defining cost functions using global versus local observables and their effects on gradient vanishing issues. These pieces of literature and theoretical insights form the foundation of our study, paving the way for us to propose new solutions and theoretical insights.

This paper presents an approach to address the barren plateaus phenomenon in quantum circuits by strategically entangling auxiliary qubits into the circuit. The gradient is then maintained while removing the auxiliary qubit. This approach transforms the circuit into a local unitary $1$-design without altering its original structure or functionality. The initial unitary operation is changed into a form similar to $\alpha I + \beta U$, which shifts the distribution from a unitary $t$-design to a unitary $(t-1)$-design. This modification preserves the anticipated gradient value while decreasing the gradient variance's reliance on the number of qubits. The methodology also involves a structured optimization process in which auxiliary qubits are gradually removed during training sessions. This assimilates the pending layers into the fixed layers with trained parameters. This preserves the circuit's core functions, ensures efficient operation, and prevents regression to a unitary $t$-design.

The significance of this study lies in its potential impact on the field of quantum computing. Our method simplifies the complexity of the quantum circuit, enhancing the efficiency of parameter training and offering a new perspective on addressing the barren plateaus challenge. This statement has important implications for developing efficient quantum algorithms and advancing quantum computing towards more practical applications. Future research could investigate the application of our method to various types of quantum circuits and parameter training techniques to optimize and expand its applicability.  
\section{\label{sec: preliminaries}PRELIMINARIES}

Our analysis employs random unitary operations along with $t$-designs. To provide a clear foundation, we will start with an introduction to these key concepts.
Let $U(N)$ be the unitary group of degree $N$, and denote the Haar measure on $U(N)$ by $\mathbf{H}$. A Haar random unitary $\mu$ is $U(N)$-valued random variable distributed according to the Haar measure $\mu\sim\mathbf{H}$ \cite{Mehta1991}.

Let $\nu$ be a probability measure on the unitary $U(N)$. A random unitary $\mu$ drawn from $\nu$ is called an $\epsilon$-approximate unitary $t$-designs if it satisfies the condition $\|\mathcal{G}_{\mu\sim\nu}^{(t)}-\mathcal{G}_{\mu\sim\mathbf{H}}^{(t)}\|_{\diamond}\leq\epsilon$, where $\|\bullet\|_{\diamond}$ represents the diamond norm, $\mathcal{G}_{\mu\sim\nu}^{(t)}$ and $\mathcal{G}_{\mu\sim\mathbf{H}}^{(t)}$ represent some specific mathematical expressions or operations involving $\mu$ \cite{Mehta1991}.

Let $X$ be a finite subset of $U(N)$, the group of $N\times N$ unitary matrices. We consider the expression
\begin{align}
\frac{1}{\abs{X}}\sum_{U\in X}f^{\otimes t}(U)=\int_{U(N)} d\mu \cdot f^{\otimes t}(U),
\end{align}
where $d\mu$ denotes the unit Haar measure on $U(N)$, satisfying $\int_{U(N)}d\mu=1$. The cardinality of $X$, denoted by $\abs{X}$, refers to the number of subsets contained within $X$.

The subset $X\subseteq U(N)$ is a $t$-design if it satisfies the condition \cite{Roy2009}:
\begin{align}
\abs{X}\geq D\left(N, \left\lceil \frac{t}{2} \right\rceil, \left\lfloor \frac{t}{2} \right\rfloor\right).
\label{eq:t_design_def}
\end{align}
where $D$ is defined in appendix~\ref{app: haar}.
Equation~\eqref{eq:t_design_def} demonstrates that unitary $t$-design does not hold universally. It necessitates the fulfillment of a specific number of conditions for its validity.

\begin{figure*}
    \centering
    \includegraphics[width=0.85\textwidth]{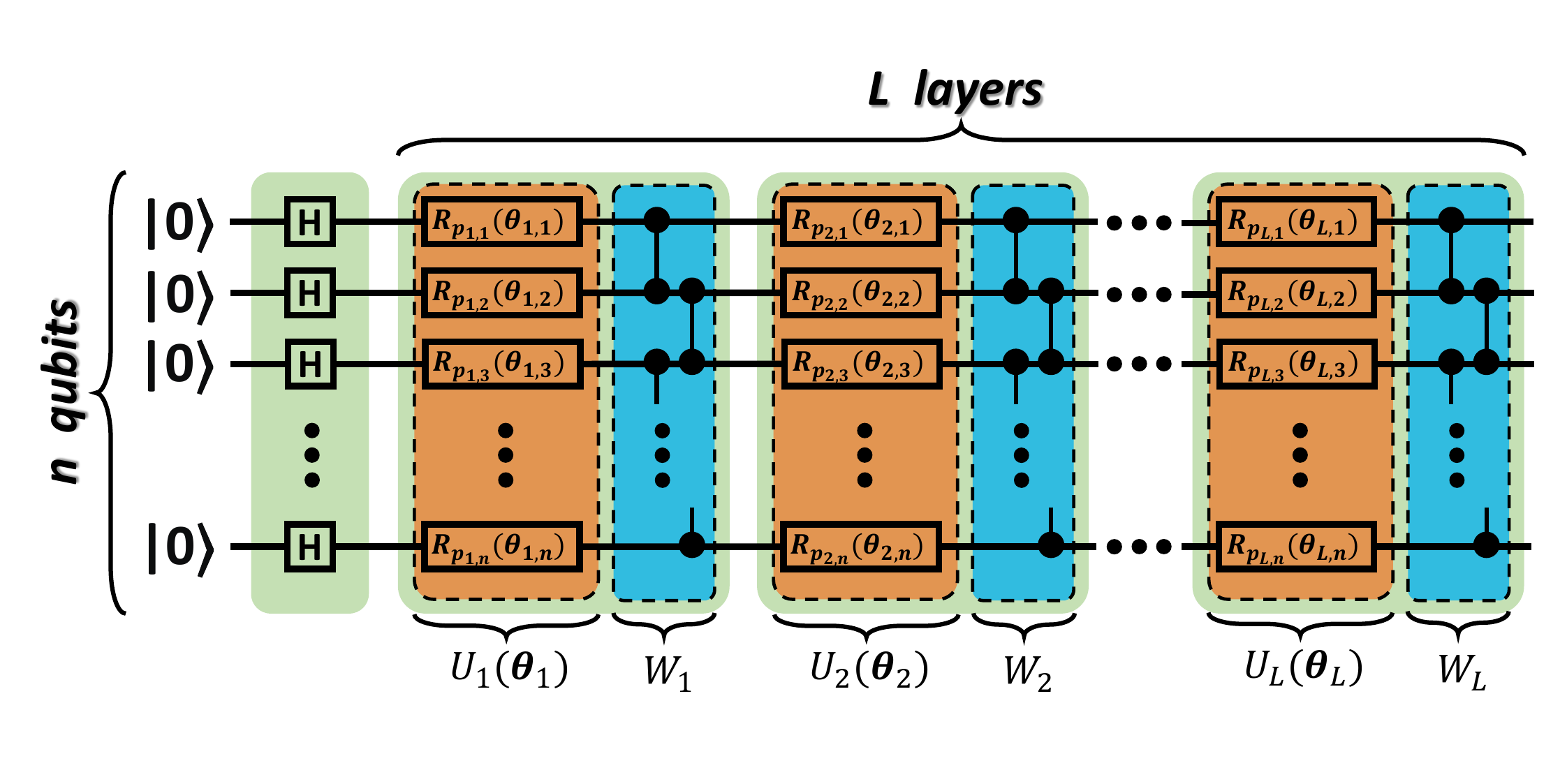}
    \caption{The architecture of random parameterized quantum circuits (RPQCs) initiates with an array of Hadamard gates, corresponding to each of the $n$ qubits, setting the groundwork for quantum superposition. This is succeeded by $L$ sequential layers, each constituting a dual-component framework. The first component of each layer, $U_l(\boldsymbol{\theta}_l)$ is parameterized, encompassing $n$ rotation gates. These gates are each defined by a set of rotational parameters $P_{l, i}\in \{X, Y, Z\}$ and a vector of angular parameters $\theta_{l, i}\in [0, 2\pi)$, which is sampled independently, enabling the modulation of quantum states in a controlled manner. The second component, $W_l$, is non-parameterized and comprises a series of CNOT gates. These gates are placed to induce entanglement between adjacent qubits.}
    \label{fig: RPQCs}
\end{figure*}

In Random Parameterized Quantum Circuits (RPQCs), comprising $n$ qubits and $L$ layers as demonstrated in Fig.~\ref{fig: RPQCs}, the unitary operator $U(\boldsymbol{\theta})$ is defined as
\begin{align}
    U(\boldsymbol{\theta})=\prod_{l=1}^{L}U_{l}(\boldsymbol{\theta}_{l})W_{l}=\prod_{l=1}^{L}\prod_{i=1}^{n}e^{-i\theta_{l,i}V_{i}}W_{l},
    \label{eq: unitary_operator}
\end{align}
where $l\in [1, L]$ is for each layer and $i\in [1, n]$ for each qubit. $U_{l}(\boldsymbol{\theta}_{l})$ and $W_{l}$ are unitary operators. Here, $V_{i}$ is a Pauli operator, and $W_{l}$ is a fixed unitary operator that does not depend on the angle $\theta_{l,i}$.

Consider a quantum circuit where the initial state is prepared in $\ket{\mathbf{init}}$. The objective function $E(\boldsymbol{\theta})$ is defined as the expectation value of a Hermitian operator $H$, which is provided externally. This expectation value is obtained by applying a unitary operation $U(\boldsymbol{\theta})$ to the initial state, and then performing a measurement corresponding to the operator $H$. The unitary operation is parameterized by a set of parameters $\boldsymbol{\theta}$. The objective function is given by:
\begin{align}
E(\boldsymbol{\theta}) = \bra{\mathbf{init}}U(\boldsymbol{\theta})^\dagger H U(\boldsymbol{\theta})\ket{\mathbf{init}},
\end{align}
where $U(\boldsymbol{\theta})^\dagger$ is the Hermitian conjugate of $U(\boldsymbol{\theta})$. 
To calculate the gradient of the objective function, we define it as follows: let $\partial_k E$ represent the $k$-th partial derivative of the function $E(\boldsymbol{\theta})$ concerning the parameter $\theta_k$, which is denoted by $\partial_k E \equiv \frac{\partial E(\boldsymbol{\theta})}{\partial\theta_k}$. It is important to note that the expected average value of this gradient is determined as
\begin{align}
\langle \partial_k E \rangle =\frac{1}{\abs{\theta_k}}\sum_{\theta_k}\partial_k E=\int d\mu \cdot \partial_k E =0.
\label{eq: bp_expactation}
\end{align}

Additionally, the variance of the gradient is given by:
\begin{align}
\mathrm{Var}[\partial_{k}E]&=\langle (\partial_k E)^2 \rangle-\langle \partial_k E \rangle^2\\
&=\frac{1}{\abs{\theta_k}}\sum_{\theta_k}(\partial_k E)^2=\int d\mu \cdot (\partial_k E)^2 \notag \\
&\approx\frac{\Tr{H^{2}}\Tr{\rho^{2}}\Tr{V_{k}^{2}}}{N^{3}-N}, \notag
\label{eq: bp_variance}
\end{align}
where $\rho=\ket{\mathbf{init}}\bra{\mathbf{init}}$.

As the number of qubits increases, the system's dimensionality grows exponentially, where $N=2^n$ and $n$ represent the number of qubits. In this scenario, the variance gradient decreases with the increase in $N$ and ultimately vanishes. This decrease in variance restricts the expressive potential of random parameterized quantum circuits \cite{McClean2018}.
 
\section{Methods}

As discussed earlier, VQAs are plagued by vanishing gradients, commonly known as the barren plateaus problem. 
The barren plateaus arise due to the circuit's unitary $2$-design characteristics. 
Previous studies \cite{Grant2019, Cerezo2021, Marrero2021} have demonstrated that adopting a local unitary $1$-design can mitigate this issue. 
This paper proposes a method to alleviate the barren plateaus problem by entanglement. It rests on modifying quantum circuits with additional qubits to transform them into a local unitary $1$-design.
Our study aims to address the barren plateaus phenomenon in quantum circuits by adding structure to the existing circuit framework. 
To achieve this, we incorporate auxiliary qubits into the system. These additional elements ensure that the original structure and function of the existing circuit remain unaltered. 
Instead, their role is primarily to mix additional information, which offers another way to address this quantum computing obstacle without necessitating structural modifications to the circuit.

\subsection{Entanglement with auxiliary qubits}

\begin{figure*}
    \centering
    \includegraphics[width=0.85\textwidth]{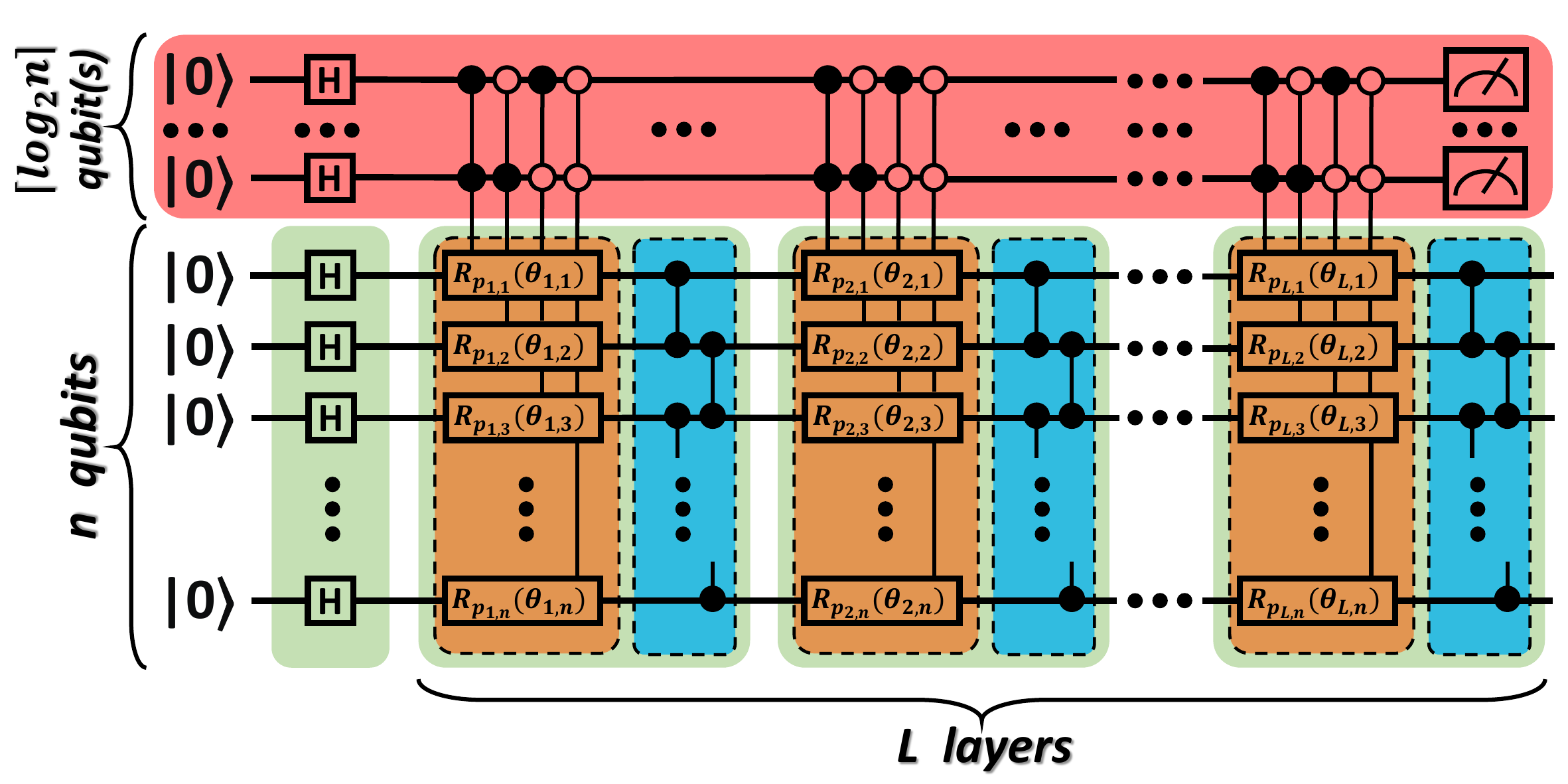}
    \caption{The structure is augmented with an additional structure involving $\lceil\log_2 n\rceil$ qubits. This supplementary segment initiates with Hadamard gates, establishing a quantum superposition state. Subsequently, these qubits control the rotation gates within the RPQCs, following a binary encoding sequence. Finally, a measurement phase is conducted on these qubits to integrate the linear combination $\alpha I + \beta U$ into the RPQCs.}
    \label{fig: structure}
\end{figure*}

In our study, the goal is to overcome the problem of barren plateaus in quantum circuits by using auxiliary qubits without altering the original circuit design.
We employ a specific configuration, as illustrated in Fig.~\ref{fig: structure}, to transform the information on the initial qubit. 
This transformation changes the form from being similar to $U$ to a new form that resembles $\alpha I + \beta U$, where $\alpha$ and $\beta$ are coefficients, and $I$ is the identity matrix. 
This approach mitigates the barren plateaus phenomenon by leveraging the additional qubits. 
This transformation introduces no new variables and changes the distribution of the entire circuit from a unitary $t$-design to a unitary $(t-1)$-design. 
While maintaining the expected value of the gradient, this transformation reduces the dependence of the variance of the gradient on the number of qubits, thereby enhancing the feasibility of circuit optimization.

Then, let $E'(\theta)$ be the expectation value of a Hermitian operator $H$ with respect to $\alpha I + \beta U$.
$E'$ and $\partial_k E'$ are represented by
\begin{align}
    E'(\mathbf{\theta})&=\bra{\mathbf{init}}(\alpha I+ \beta U(\mathbf{\theta}))^\dagger H(\alpha I+ \beta U(\mathbf{\theta}))\ket{\mathbf{init}},\\
    \partial_k E'&=i\alpha\beta\bra{\mathbf{init}}[R, H]\ket{\mathbf{init}}+\beta^2(\partial_k E),
\end{align}
where $R=U_+V_kU_-$. When calculating the expectation of $\partial_k E'$ with respect to $U$, the first and the second moments are
\begin{align}
    \langle \partial_k E' \rangle &=0,\\
    \langle (\partial_k E')^2 \rangle &=\frac{2\alpha^2 \beta^2}{N}\Tr{H^2\rho}\Tr{\rho}+\beta^4\langle (\partial_k E)^2 \rangle,
    \label{eq: unitary_1_design}
\end{align}
where $\left\langle\left(\partial_k E^{\prime}\right)^2\right\rangle$ follows from the property of a unitary $1$-design.
By observing the unitary $1$-design part of $\langle (\partial_k E')^2 \rangle$, we can find that
\begin{align}
\frac{2\alpha^{2}\beta^{2}}{N}\Tr{H^{2}\rho}\Tr{\rho}\leq\frac{1}{2N}\Tr{H^{2}\rho}\Tr{\rho}.
\end{align}
The maximum value is attained when $\alpha=\beta=\frac{1}{\sqrt{2}}$.

\subsection{Eliminate auxiliary qubits}
In certain scenarios, using the original circuit becomes an unavoidable choice. However, when applying this optimization technique to adjust the parameters of these original circuits, it often becomes evident that the optimized parameters are not equivalent to those in the original circuit. This discrepancy arises from using different assumptions made during the optimization process.

\begin{figure*}
    \centering
    \includegraphics[width=0.85\textwidth]{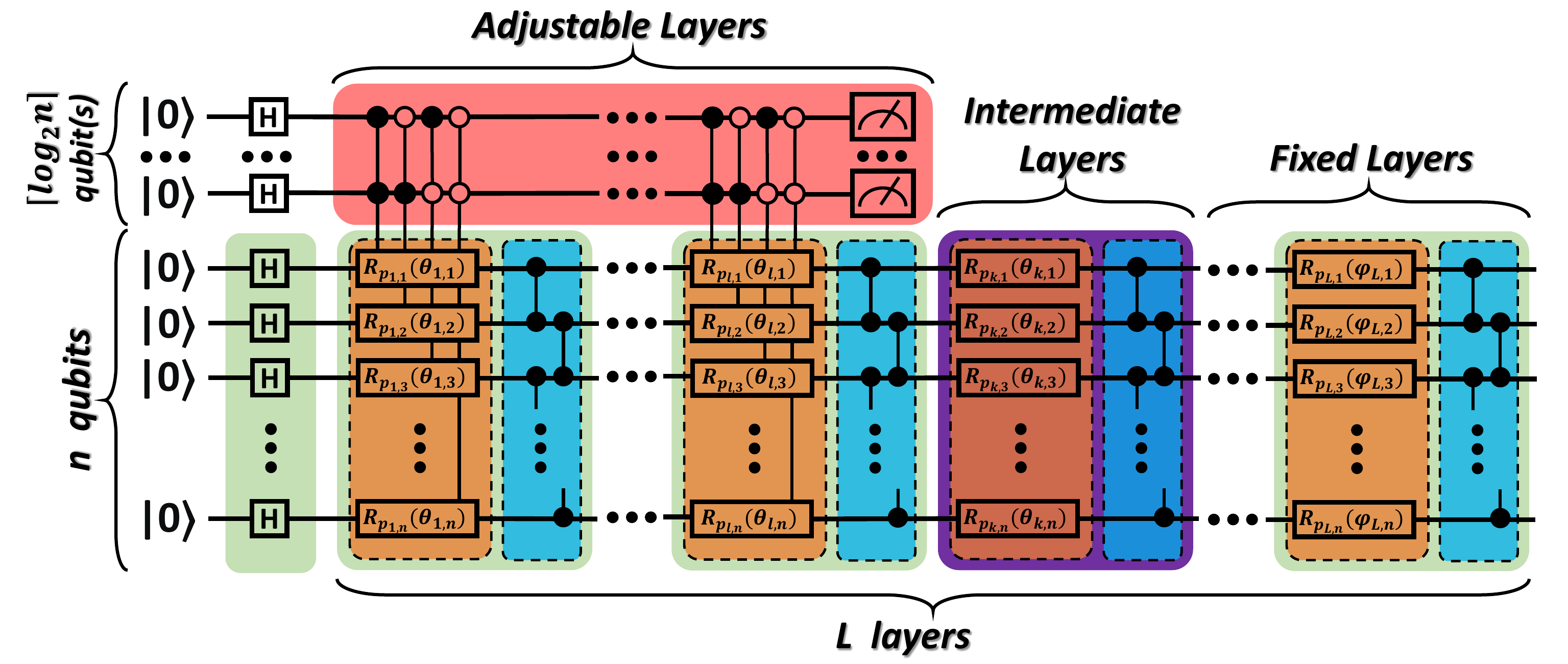}
    \caption{The structure consists of three sections: adjustable layers, intermediate layers, and fixed layers. Within this structure, the parameters in the adjustable and intermediate layers are updated simultaneously. During the update process, once the intermediate layer parameters are determined, this section becomes fixed layers with unchanging parameters. At this point, the last part of the adjustable layers becomes the new intermediate layers. This entire process is repeated continuously until the length of fixed layers equals $L$, at which point the optimization is complete.}
    \label{fig: proposed}
\end{figure*}

The original circuit aimed to minimize the energy $E(\boldsymbol{\theta}) = \bra{\mathbf{init}}U(\boldsymbol{\theta})^\dagger H U(\boldsymbol{\theta})\ket{\mathbf{init}}$. However, with the introduction of auxiliary qubits, the energy changed to $E'(\mathbf{\theta})=\bra{\mathbf{init}}(\alpha I+ \beta U(\mathbf{\theta}))^\dagger H(\alpha I+ \beta U(\mathbf{\theta}))\ket{\mathbf{init}}$. Therefore, parameter sets that minimize $E(\boldsymbol{\theta})$ and $E'(\boldsymbol{\theta})$ are generally different. Therefore, we will employ the structure depicted in Fig.~\ref{fig: proposed}.  This involves gradually eliminating the auxiliary qubits through multiple training sessions until the circuit returns to its original configuration.

By successively transforming the adjustable layers into fixed layers through the intermediate layer, while keeping the parameters within the fixed layers unchanged, we can gradually restore the circuit to its original structure and eventually eliminate all auxiliary qubits. Adjustable and intermediate layers must be updated simultaneously during this training process. In this process, the adjustable layers are unitary $1$-designs, and the intermediate layers are unitary $2$-designs. To generate better gradient values, we must ensure that the number of layers in the layer to be processed is small enough for optimal results. This approach effectively overcomes the challenge posed by the barren plateaus phenomenon and ensures that parameter optimization continues smoothly.

The parameterized layers of the unitary operator in Fig.~\ref{fig: proposed} conform to a unitary $1$-design. We assume that $\boldsymbol{\varphi}$ represents a set of fixed parameters exempt from training to substantiate this claim. 
Let $E''$ be the energy $E'$ with specific parameter values $\alpha = \beta = 1/\sqrt{2}$.
Consequently, we can demonstrate the following results for $E''$ and its corresponding gradient $\partial_k E''$.
\begin{align}
    E''(\boldsymbol{\theta}) &= \Tr{\frac{\rho}{2}(I+U(\boldsymbol{\theta}))^\dagger U'(\boldsymbol{\varphi})^\dagger H U'(\boldsymbol{\varphi})(I+U(\boldsymbol{\theta}))},\\
    \partial_k E''(\boldsymbol{\theta}) &= \frac{i}{2}\Tr{\rho[R,H_\varphi]+H_{\varphi,+}[\rho_-, V_k]},
\end{align}
where $H_\varphi=U'(\boldsymbol{\varphi})^\dagger H U'(\boldsymbol{\varphi})$, $\rho_-=U_-^\dagger \rho U_-$ and $H_{\varphi,+}=U_+^\dagger H_\varphi U_+$. Then, if $U_+$ is at least a unitary $1$-design, the expectation and variance of $\partial_k E''$ are
\begin{align}
    \langle \partial_k E'' \rangle &=0,\\
    \mathrm{Var}[\partial_k E''] &=\frac{1}{2N} \Tr{\rho H_\varphi^2}\Tr{\rho}+O(N^{-1}).
\end{align}
As the number of qubits increases, the variance of $\partial_k E''$ will be larger than the variance of $\partial_k E$. This results in the distribution of $\partial_k E$ becoming more concentrated around its mean value of zero compared to the distribution of $\partial_k E''$.
Thus, the suggested approach is expected to achieve the effective value more reliably than the unitary $2$-design structure, which helps overcome the barren plateaus issue.

\section{Experiment}

In order to demonstrate the effectiveness of this proposed method, we perform a numerical experiment.
The purpose of this experiment is to compare the performance of three quantum circuit models by using Pennylane \cite{bergholm2018}.
The evaluated models are standard RPQCs, which serve as the unitary $2$-design baseline model, a unitary $1$-design structure incorporating auxiliary qubits, and the proposed structure that optimizes by eliminating these auxiliary qubits. The experimental design involves sampling 100 quantum circuits that are randomly generated according to each model's configuration. For each circuit, we compute the variance of its gradient for the target operator $H = Z_1 \otimes Z_2$, where $H$ is a single Pauli ZZ operator acting on the first and second qubits \cite{Cerezo2021}. This comparison aims to reveal the impact of different design choices on the performance of quantum circuits, particularly in terms of gradient variance.

\begin{figure}[h]
    \centering
    \includegraphics[width=\linewidth]{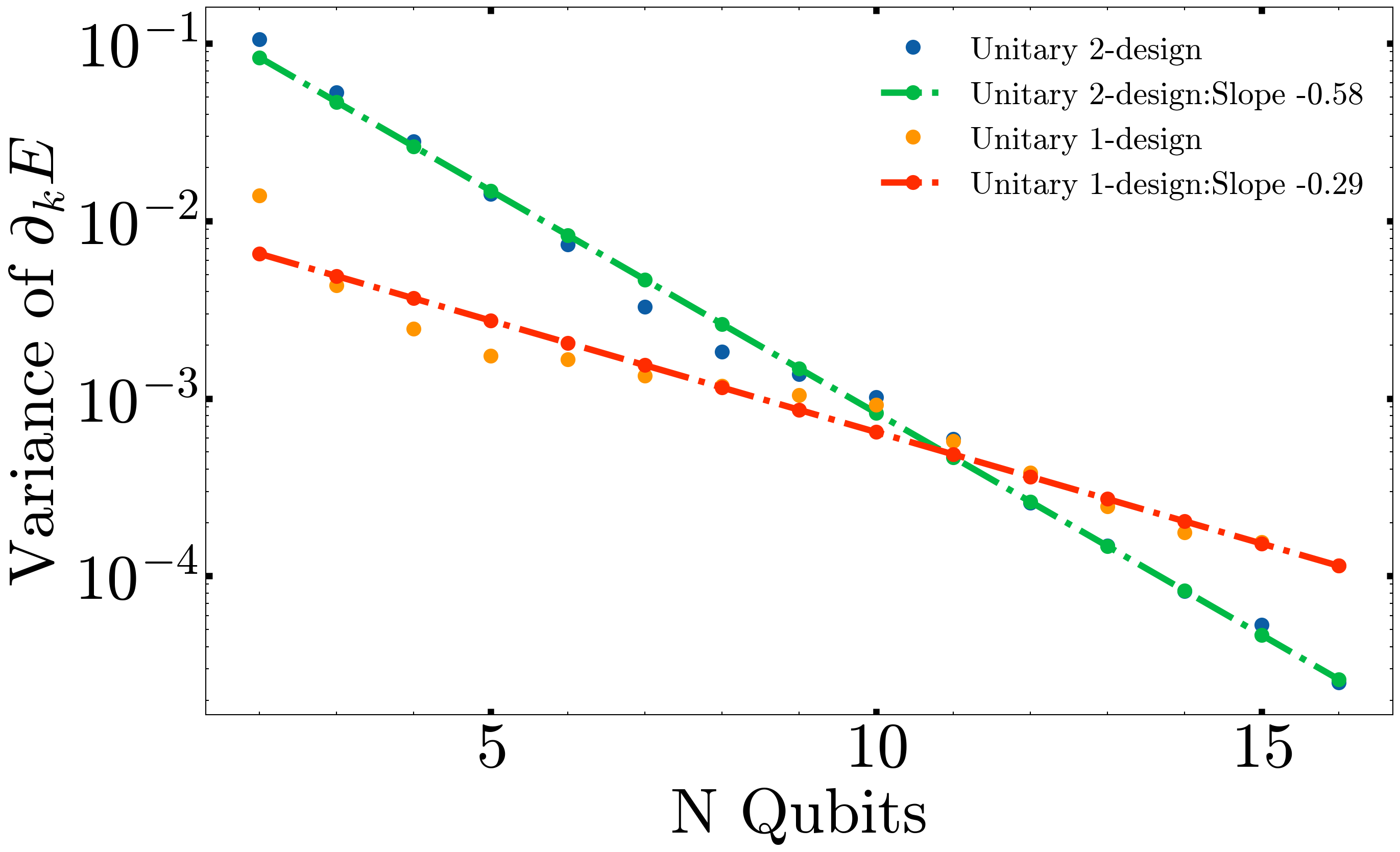}
    \caption{The variances in gradients were compared with different numbers of qubits under unitary t-designs of different orders. The blue and orange points represent the variances for t=2 and t=1, respectively, across a range of qubit numbers. The variances are approximated by polynomial functions represented by the green and red lines, and their slopes are also given.}
    \label{fig: exp_1}
\end{figure}

Firstly, we are concerned about the impact of the number of qubits on the variance of $\partial_k E$. We configure all structures to have 100 layers, with the number of qubits ranging from 2 to 16. 
The experimental results demonstrate that the variance of quantum circuits with a unitary $2$-design structure decreases exponentially as the number of qubits increases. The slope of the curve is approximately -0.58, as shown in Fig.~\ref{fig: exp_1}. This trend suggests that as the number of qubits increases, the gradient of the entire quantum circuit becomes easier to zero out, resulting in a significant decline in performance. The variance of quantum circuits with a unitary $1$-design structure also decreases exponentially with the number of qubits. However, the slope of this decrease is only about half that of the unitary $2$-design structure, which is consistent with the Eq.~\eqref{eq: unitary_1_design}.

Subsequently, we evaluate the performance of the gradient to $\theta_{1,1}$, the first parameter in the quantum circuit in Fig.~\ref{fig: proposed}, across varying configurations of qubits and layers within the quantum circuit architecture. Initially, the structure adheres to a unitary $2$-design structure. This structure transforms a unitary $1$-design upon entanglement with auxiliary qubits. After setting the pending layers to 20 and following the removal of these auxiliary qubits, the structure transitions into the proposed structure in this study. Our experimental assessment encompasses a range of qubits from 2 to 16 and layers from 5 to 500, providing a comprehensive overview of the impact of these variables on the gradient of the quantum circuit's performance.

\begin{figure}[h]
    \centering
    \includegraphics[width=\linewidth]{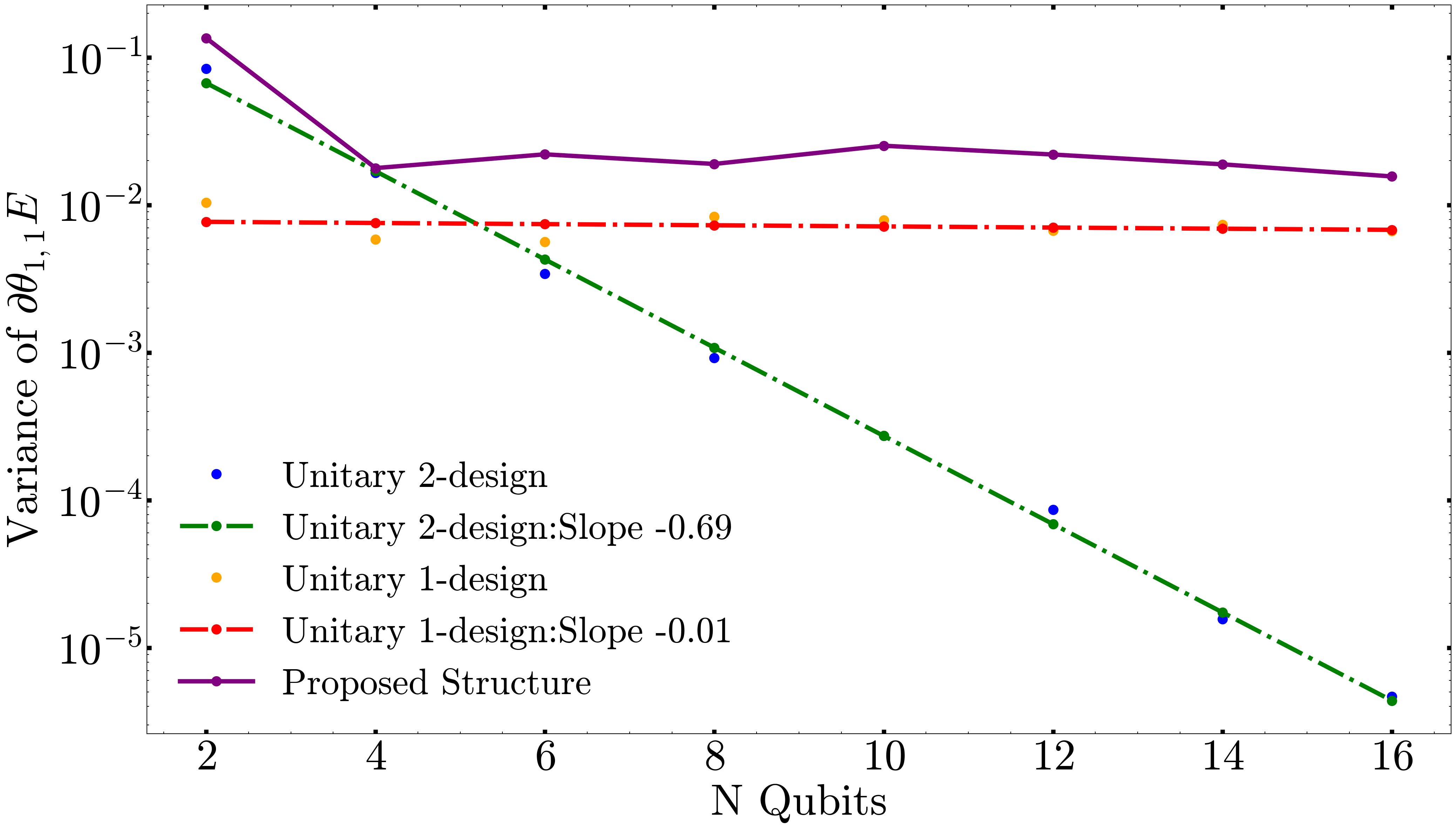}
    \caption{The variances in gradients were compared with different numbers of qubits under different structures. The blue and orange points represent the variances for unitary $2$-design and $1$-design, respectively, across a range of qubit numbers. The variances are approximated by polynomial functions represented by the green and red lines. The slopes of these lines are also given. The variances associated with the proposed structure are illustrated via purple lines.}
    \label{fig: exp_2_1}
\end{figure}

Consistent with previous findings in Fig.~\ref{fig: exp_1}, the influence of the number of qubits on the variance of a single parameter demonstrates a similar pattern. The variance of quantum circuits with a unitary $2$-design structure exhibits an exponential decline as the number of qubits increases, with a slope of approximately -0.69, as illustrated in Fig.~\ref{fig: exp_2_1}. This trend highlights that as the number of qubits increases, the quantum circuit's gradient becomes more susceptible to nullification on a local variable scale. This indicates the barren plateaus phenomenon, which affects the global landscape and local variational parameters, resulting in a significant performance deterioration. Conversely, quantum circuits with a unitary $1$-design structure display a variance that shows no significant correlation with the number of qubits, with the slope nearing zero. The proposed method inherits the characteristics of the unitary $1$-design and maintains consistent variance as the number of qubits increases.

\begin{figure}[h]
    \centering
    \includegraphics[width=\linewidth]{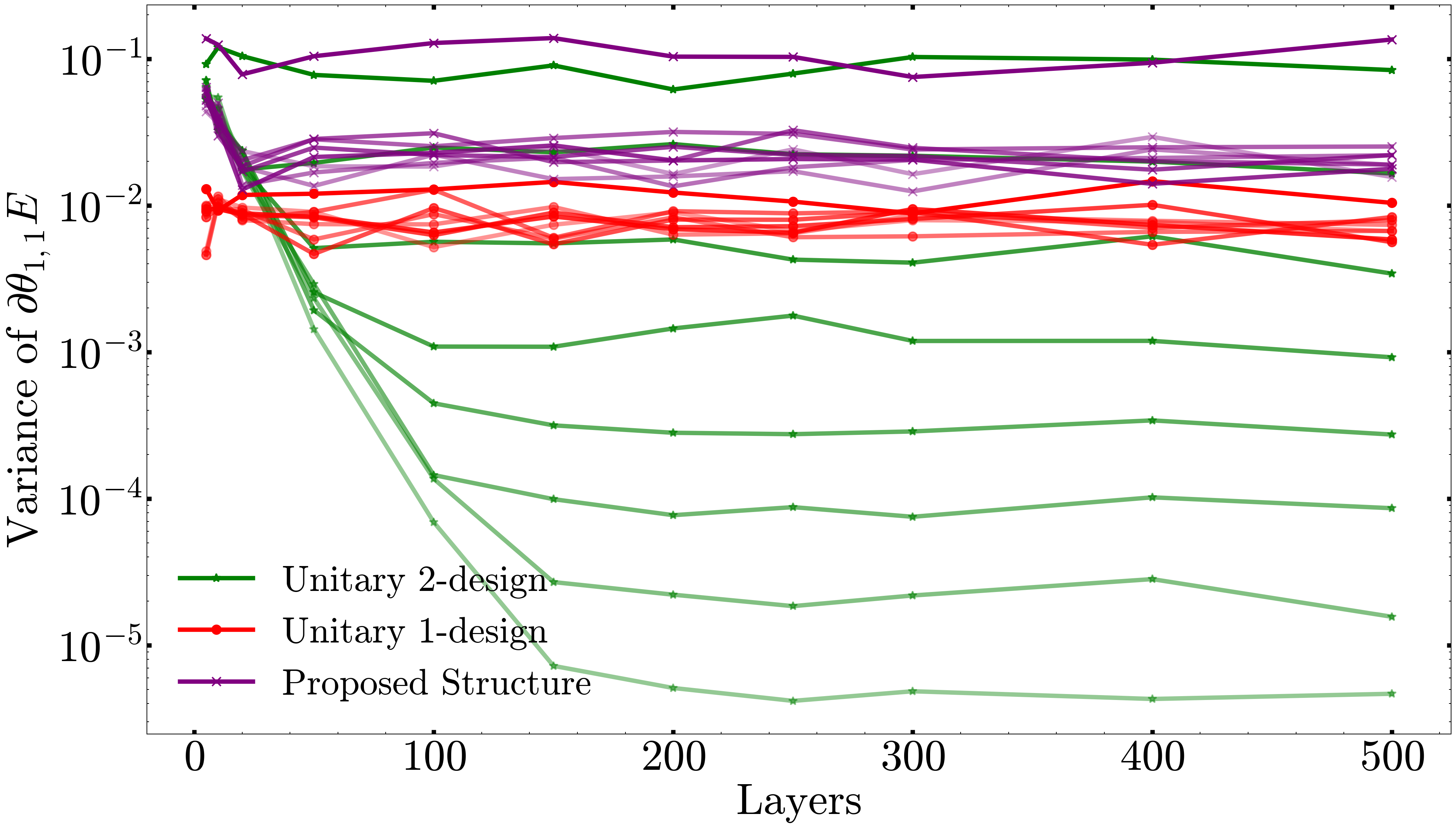}
    \caption{The variance of the gradient is compared for different numbers of layers under unitary t-designs of varying orders and proposed structure. The variances for t=2 and t=1 designs are depicted by the green and red lines, respectively, while the purple lines represent the proposed structure. The shading transitions from dark to light are quantum circuits with even numbers of qubits from 2- to 16-qubit, where the top line represents 2 qubits.}
    \label{fig: exp_2_2}
\end{figure}

Regarding the impact of the number of layers on variance, Fig.~\ref{fig: exp_2_2} shows that the variance decreases as the number of layers increases for quantum circuits with a unitary $2$-design structure. Additionally, the decrease becomes more pronounced as the number of qubits increases. There is a smaller range of variance fluctuation for quantum circuits with a unitary $1$-design structure, and the number of layers is not significantly affected. However, it should be noted that under conditions of fewer qubits or shallower layers, the unitary $2$-design structure tends to have an advantage. To utilize both advantages, the proposed structure follows a unitary $2$-design when the pending layers are small. As the layers deepen, the structure transitions to a unitary $1$-design and is more likely to achieve an effective gradient.

\begin{figure}[h]
    \centering
    \includegraphics[width=\linewidth]{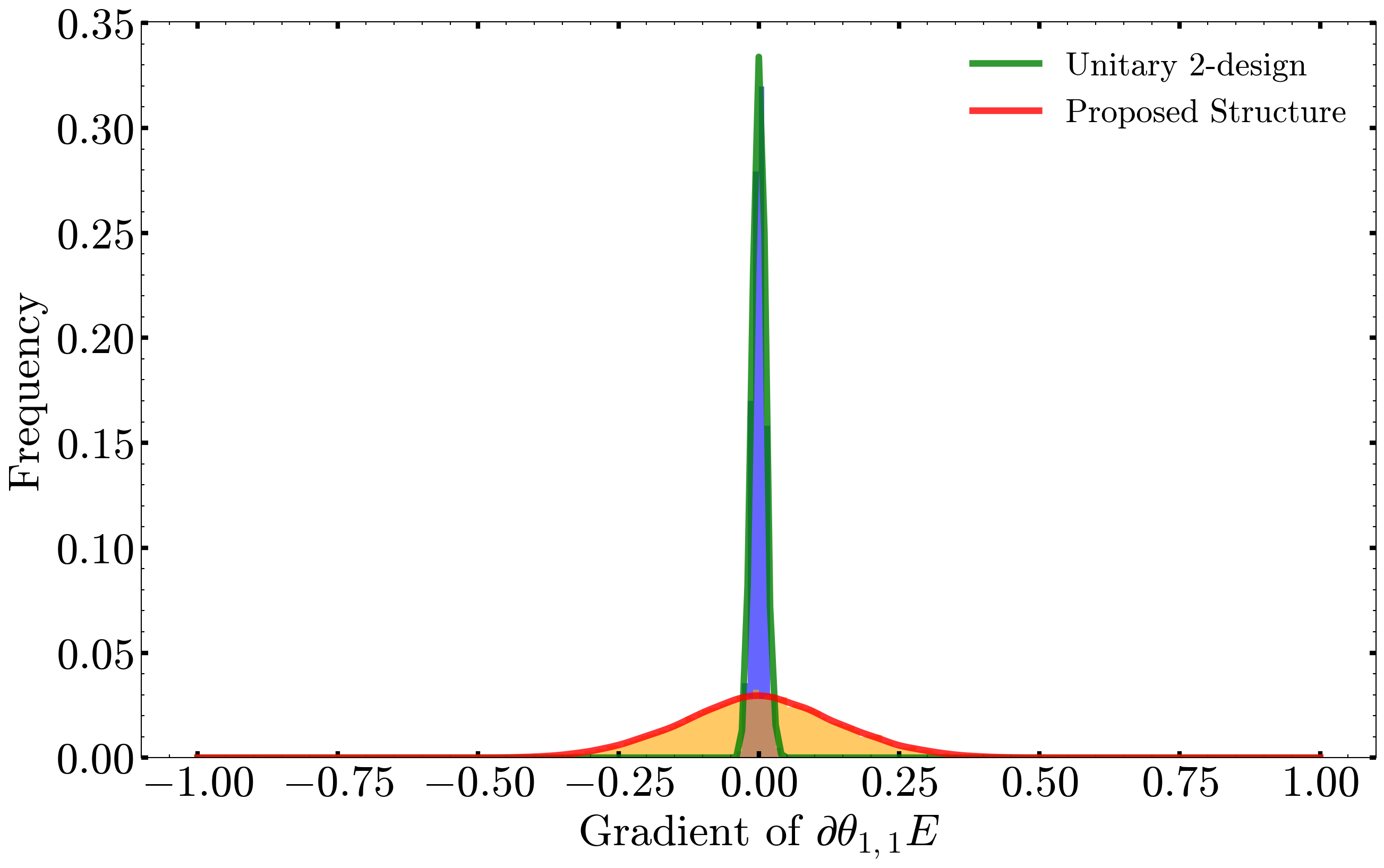}
    \caption{Gradient Distribution $\partial\theta_{1,1}E$ in a 10-Qubit, 500-Layer Quantum Circuit. The orange and blue histograms depict the frequency of gradient values for the proposed and the unitary $2$-design structure, respectively. The red and green curves represent Gaussian kernel-density estimates fitted to the histograms, capturing the distribution trends for each structure.}
    \label{fig: exp_2_3}
\end{figure}

Then we investigate the gradient distribution $\partial\theta_{1,1}E$ within a quantum circuit composed of 10 qubits and 500 layers in in Fig.~\ref{fig: exp_2_3}. The distribution profile for the proposed structure and a unitary $2$-design are assessed for their respective potentials in gradient optimization. The unitary $2$-design structure exhibits a steep, narrow distribution of gradient values, closely centralized near zero and conforming to a normal distribution expressed as $\mathcal{N}(0, 2.738\times 10^{-4})$, where $\mathcal{N}$ denotes the normal distribution with a mean of $0$ and a variance of $2.738\times 10^{-4}$. This highlights the barren plateaus phenomenon, where optimization becomes challenging due to vanishing gradients. In contrast, the proposed structure demonstrates a substantially wider distribution of gradient values, conforming to a normal distribution expressed as $\mathcal{N}(0, 1.655\times 10^{-2})$, indicating a decreased propensity for gradients to converge to zero. This increases the likelihood of locating non-trivial gradient values conducive to effective optimization. The proposed architecture outperforms the other structure by avoiding barren plateaus facilitating more robust quantum circuit training and optimization.

\begin{figure*}[h]
    \centering
    \includegraphics[width=\textwidth]{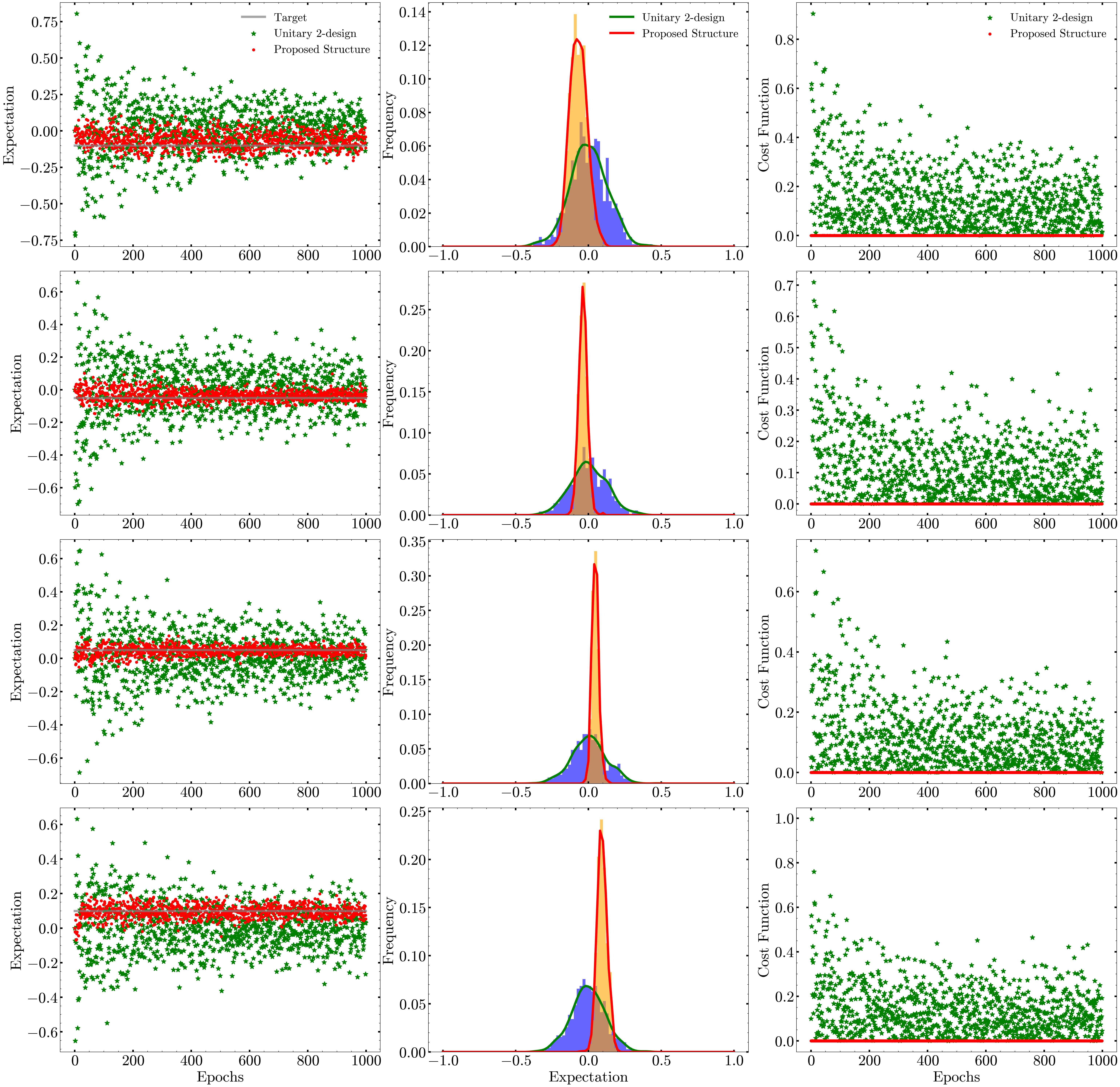}
    \caption{Comparative Analysis of 10-qubit and 100-layer circuits across four experiments, targeting values at -0.1, -0.05, 0.05, and 0.1. Each row represents one experiment, showcasing epoch-wise expectation results (left), the frequency distribution of expectation (middle), and epoch-cost function value (right). In Epoch-Expectation, the horizontal grey line denotes the target, with red points for the proposed method's outputs and green points for the unitary $2$-design structure's outputs. In Expectation-Frequency, blue and yellow bars depict the occurrence frequency of the proposed method and the unitary $2$-design structure's expectations, respectively, with green and red curves representing the Gaussian kernel-density estimate fit distribution. In the Epoch-Cost Function, red points show the proposed method's cost values, while green points are for the unitary $2$-design structure's cost values.}
    \label{fig: exp_3}
\end{figure*}

Finally, we conduct four separate experiments in Fig.~\ref{fig: exp_3} to evaluate the effectiveness of the unitary $2$-design and proposed structure methodologies in achieving fixed target values of -0.1, -0.05, 0.05, and 0.1 in expectation and cost function. Each experiment is configured with a 10-qubit, 100-layer structure. The proposed structure uses a pending layer setting of 1 and is trained for 10 epochs per pending structure, totaling 1000 epochs. The cost function is $(\mathrm{expectation}-\mathrm{target})^2$ and the optimizer is \textit{pennylane.AdamOptimizer()} \cite{bergholm2018}.

Over the 1000 observed epochs for each experiment in Fig.~\ref{fig: exp_3}, the proposed structure's expectations tend to cluster closely around the target line, unlike the unitary $2$-design, whose data points are notably more dispersed. The proposed structure's convergence towards the target value indicates superior performance, highlighting its effectiveness over the unitary $2$-design in achieving the desired outcomes.

Calculating the probability distribution of expected values across 300 to 1000 epochs allows us to determine the variability of the predicted value. A narrower peak indicates a more consistent approach to the target, while a wider distribution indicates a greater variance. The central peak of each distribution corresponds to the most frequently occurring expected value within 1000 periods. Ideally, these peaks should be the targets. Compared to the unitary $2$-design structure, the proposed structure's distribution around the target value is sharper and has larger peaks. This indicates that the proposed structure is closer to the target's expected value.

The cost function is a measure used to evaluate the model's performance, with lower values indicating better performance. The proposed structure is more likely to obtain effective gradients during training, allowing its cost function maintain at a level less than $10^{-4}$. In contrast, the unitary $2$-design structure cannot achieve this, preventing the cost function from converging to a smaller value.

\section{Conclusion}

This paper addressed a central issue in quantum computing – the barren plateaus phenomenon. This phenomenon presents a significant challenge in quantum machine learning and optimization algorithms. Large-scale quantum circuits are characterized by a vanishing gradient variance, which must be addressed. Our approach successfully transforms the quantum circuit from a unitary $2$-design to a unitary $1$-design without changing the original structure, marking a significant stride in gradient optimization. We introduced auxiliary control qubits that can be eliminated to achieve this transformation.

Our experiments systematically demonstrated the advantages of the proposed structure over global unitary $2$-design quantum circuits. The entanglement and elimination of auxiliary qubits facilitate the gradual transition from a unitary $2$-design to a unitary $1$-design. This approach effectively mitigates the challenges posed by barren plateaus. The experimental data consistently indicated that the proposed structure maintains a stable gradient variance, avoiding the exponential decline associated with increased qubits. This is a notable issue in global unitary $2$-design structures. Moreover, the proposed structure showcases a diminished sensitivity to the number of layers, maintaining effective gradients more reliably than the unitary $2$-design, particularly in circuits with more layers.

The analysis of gradient distributions emphasizes the superiority of the proposed structure. The unitary $2$-design is prone to narrow gradient distributions, which centralize near zero and indicate barren plateaus. In contrast, the proposed structure exhibits a broader gradient distribution, increasing the probability of achieving non-trivial gradients and successful optimization.

In achieving fixed target values for the expectation and cost function, the proposed structure demonstrates a pronounced ability to align with target values closely. It outperforms the unitary $2$-design in both consistency and cost function minimization. This is evidenced by the sharper and taller peaks in the probability distribution of expected values and the ability to maintain the cost function consistently low throughout the training process.

Our method has demonstrated significant efficacy in random parameterized quantum circuits (RPQCs). We have experimentally validated our approach to RPQCs, demonstrating its theoretical feasibility and practical applicability. Importantly, our method achieves these benefits without introducing additional variables or significantly altering the existing quantum circuit structure. This substantially reduces implementation complexity and cost in practical applications.

The methodology and results of this study may pave new pathways in the field of quantum computing. Our study will be extended to other types of quantum circuits, such as quantum neural networks (QNNs). Applying the same method in these circuits is expected to yield positive results in gradient optimization. This strategy effectively addresses the barren plateaus problem and opens up new opportunities for designing and optimizing future quantum algorithms.

In summary, this study proposes an innovative and practical method that effectively solves an essential issue in quantum computing. Our study aims to enhance the comprehension and optimization of gradient behaviors in large-scale quantum circuits. It provides valuable guidance for future gradient optimization strategies in quantum machine learning and other advanced quantum computing applications.

The code used for this work is released in \cite{code}. 
\nocite{*}

%

\appendix
\onecolumngrid
\section{Haar Measure and Weingarten Function}
\label{app: haar}

The unitary representation is introduced in this section. \cite{Roy2009} presents the following theorem. Let $U(N)$ denote the unitary group of degree $N$. The irreducible representations of $U(N)$ which occur in $(\mathbb{C}^N)^{\otimes r}\otimes(\mathbb{C}^{N*})^{\otimes s}$ are precisely those indexed by non-increasing, length-$N$ integer sequences $\mu=(\mu_1, \mu_2, \dots, \mu_N)$, under the conditions:
\begin{enumerate}
    \item The number of the elements: $\abs{\mu}=r-s$
    \item The number of the positive elements: $\abs{\mu_+}\leq r$
\end{enumerate}

Furthermore, the dimension of each irreducible representation indexed by such a sequence $\mu$ is given by:
\begin{align}
    d_\mu=\prod_{1\leq i\leq j\leq N}\frac{\mu_i-\mu_j+j-i}{j-i}.
\end{align}
For positive integers $N$, $r$, and $s$, the total dimension, denoted as $D(N,r,s)$, contributed by these representations to the tensor product space is calculated by the square sum of the dimensions of all such irreducible representations that satisfy the conditions above:
\begin{align}
    D(N,r,s):= \sum_{\substack{\abs{\mu}=r-s \\ \abs{\mu_+}\leq r }} d_{\mu}^{2}.
\end{align}

Then think about a special haar measure \cite{collins2003}: Let $N$ be a positive integer and $\Vec{i}=(1, 2, \dots, i_p)$, $\Vec{j}=(1, 2, \dots, i_q)$ be tuples of positive integers from $(1, 2, \dots, N)$. Then, 
\begin{align}
    I_{N,p,q} &= \int_{U(N)} d\mu \cdot U_{i_1j_1}U_{i_2j_2}\dots U_{i_pj_p}U_{i'_1j'_1}^\ast \dots U_{i'_qj'_q}^\ast =
    \begin{cases} 
        0, & \text{if } p \neq q \\
        \sum_{\sigma, \tau\in S_p}\delta_{(\sigma, \tau)}Wg(N, \sigma\tau^{-1}), & \text{if } p = q 
    \end{cases}
\label{eq: weingarten}
\end{align}

In this function, $\delta_{(\sigma, \tau)}=\delta_{i_1 i'_{\sigma(1)}} \cdots \delta_{i_q i'_{\sigma(q)}} \delta_{j_1 j'_{\tau(1)}} \cdots \delta_{j_q j'_{\tau(q)}}$ and $Wg$ is the Weingarten function \cite{weingarten1978}, given by

\begin{align}
    Wg(N, \sigma) = \frac{1}{q!^2} \sum_{\lambda} \frac{\chi^{\lambda}(1)^2 \chi^{\lambda}(\sigma)}{s_{\lambda, N}(1)},
\end{align}
where the sum over all partitions $\lambda$ of $q$. The character corresponding to the partition $\lambda$ is represented by $\chi^\lambda$, and $s$ is the Schur polynomial of $\lambda$. Therefore, $s_{\lambda, N}(1)$ represents the dimension of the representation of $U(N)$ corresponding to $\lambda$.

Then, we require certain conclusions about the Haar measure, which can be proved by Equation~\eqref{eq: weingarten}. First, when $U$ is a unitary $1$-design,  the following equation holds due to the result that $Wg(N,(1))=\frac{1}{N}$.
\begin{align}
    \Tr{\int d\mu\cdot U^\dagger AUB} = \frac{\Tr{A}\Tr{B}}{N}.
    \label{eq: haar_1}
\end{align}

Similarly, Equation~\eqref{eq: haar_2} is valid when $U$ represents a unitary $2$-design, with $Wg(N,(1,1))=\frac{1}{N^2-1}$ and $Wg(N,(2))=-\frac{1}{N^3-N}$.
\begin{align}
    \label{eq: haar_2}
    &\Tr{\int d\mu\cdot U^\dagger AUBU^\dagger CUD}\\
    =&\frac{\Tr{A}\Tr{C}\Tr{BD}+\Tr{AC}\Tr{B}\Tr{D}}{N^2-1}
    -\frac{\Tr{A}\Tr{B}\Tr{C}\Tr{D}+\Tr{AC}\Tr{BD}}{N^3-N}. \notag
\end{align}

\section{Detail of Entanglement with Auxiliary Qubits}
\label{app: entanglement}

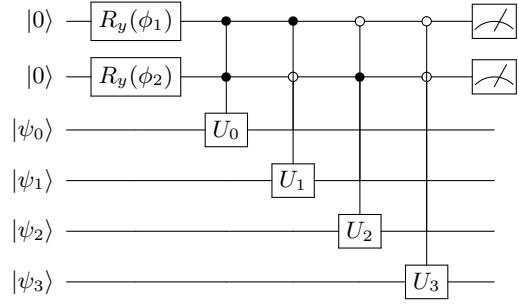
\begin{figure}[h]
    \centering
    \[
    \Qcircuit @C=1em @R=.7em {
    \lstick{\ket{0}}    & \gate{R_y(\phi_1)} & \ctrl{1} & \ctrl{2} & \ctrlo{3} & \ctrlo{4} & \meter \\
    \lstick{\ket{0}}    & \gate{R_y(\phi_2)} & \ctrl{1} & \ctrlo{2} & \ctrl{3} & \ctrlo{4} & \meter \\
    \lstick{\ket{\psi_0}} & \qw & \gate{U_0} & \qw & \qw & \qw & \qw\\
    \lstick{\ket{\psi_1}} & \qw & \qw & \gate{U_1} & \qw & \qw & \qw\\
    \lstick{\ket{\psi_2}} & \qw & \qw & \qw & \gate{U_2} & \qw & \qw\\
    \lstick{\ket{\psi_3}} & \qw & \qw & \qw & \qw & \gate{U_3} & \qw\\
    }
    \]
    \caption{This simplified figure illustrates how to entangle an $n$-qubit single-layer configuration with $\lceil \log_2n\rceil$ auxiliary qubits for a parameterized layer. The unitary gate $U_i$ represents the rotation gate in RPQCs.}
    \label{fig: smplified_example}
\end{figure}

To ensure clarity and focus, we present the circuit structure of an $n$-qubit RPQCs entangled with $\lceil \log_2n\rceil$-qubit auxiliary qubits as a single-layer configuration for a parameterized layer, shown in Fig.~\ref{fig: smplified_example}. 

Before measurement, the unitary operator of this circuit appears as follows:
\begin{align}
U = & \prod_{i=1}^{n} [\rho_{\Bar{i}} \otimes I+\rho_{i} \otimes U_{i}]_i \otimes I_{\Bar{i}}^{\otimes n-1} \cdot \left[ \bigotimes_{i=1}^{\lceil \log_2n\rceil} R_y(\phi_i)\otimes I^{\otimes n} \right],
\end{align}
where $I_{\Bar{i}}^{\otimes n-1}$ is the identity matrix on all qubits except qubit $i$, $\rho_i=\ket{i}\bra{i}$ and $\rho_{\Bar{i}}=I^{\otimes \lceil \log_2n\rceil}-\rho_i$. After measurement, the unitary operator of the parameterized layer is the partial trace for auxiliary qubits, the operator will transfer to
\begin{align}
\Tr_{a}\{U\}
= & \prod_{i=1}^{n} [\Tr{R_y(\phi_i)\rho_{\Bar{i}}} \otimes I+\Tr{R_y(\phi_i)\rho_{i}} \otimes U_i]_i \otimes I_{\Bar{i}}^{\otimes n-1} \\
= & \prod_{i=1}^{n} [\alpha_i I+\beta_i U_{i}]_i \otimes I_{\Bar{i}}^{\otimes n-1} \notag \\
= & \bigotimes_{i=1}^{n}\alpha_i I+\beta_i U_i. \notag
\end{align}

Then we prove the conclusions about $E'$. A parameterized quantum circuit can be characterized by a sequential application of unitary operations $U(\boldsymbol{\theta})$ is defined as
\begin{align}
    U(\boldsymbol{\theta}) = \prod_{l=1}^{L} U_{l}(\mathbf{\theta_{l}})W_{l} = \prod_{l=1}^{L} \prod_{i=1}^{n}e^{-i\theta_{i,l}V_{i}}W_{l},
\end{align}
where $U_{l}(\theta_{l})$ and $W_{l}$ are unitary operators. Here, $V_{i}$ is a Pauli operator, and $W_{l}$ is a fixed unitary operator that does not depend on the angle $\theta_{i,l}$. If we calculate the gradient of $U$, we can get the result that
\begin{align}
    \label{eq: partial_u}
    \partial_k U(\boldsymbol{\theta}) = \prod_{l=k}^{L} U_{l}(\mathbf{\theta_{l}})W_{l}(-iV_k)\prod_{l=1}^{k-1} U_{l}(\mathbf{\theta_{l}})W_{l} = U_+(-iV_k)U_-=-iR.
\end{align}

Then, we utilize the structure with auxiliary qubits in Fig.~\ref{fig: smplified_example}, which enables us to derive the objective function as follows
\begin{align}
    E'(\boldsymbol{\theta})&=\bra{0}(\alpha I+ \beta U(\boldsymbol{\theta}))^\dagger H(\alpha I+ \beta U(\boldsymbol{\theta}))\ket{0}\\
    &=\alpha^2\bra{0}H\ket{0}+\alpha\beta[\bra{0}HU\ket{0}+\bra{0}U^\dagger H\ket{0}]+\beta^2 E. \notag
\end{align}

Next, we compute the gradient of $E'(\mathbf{\theta})$, and the first component is $0$. The remaining part can be calculated below using Equation~\eqref{eq: partial_u}.
\begin{align}
\partial_k E'&=\alpha\beta(-i\bra{0}HR\ket{0}+i\bra{0}R^\dagger H\ket{0})+\beta^2(\partial_k E) \\
&=i\alpha\beta\bra{0}[R, H]\ket{0}+\beta^2(\partial_k E). \notag
\end{align}

Given that $p \neq q$ in Equation~\eqref{eq: weingarten} and considering the condition $\langle \partial_k E\rangle=0$, the expectation of the gradient is zero.

Then, we need to calculate the variance. Because of the function that
\begin{align}
    \label{eq: var}
    \mathrm{Var}[\partial_k E']&=\langle(\partial_k E')^2\rangle-\langle\partial_k E'\rangle^2=\langle(\partial_k E')^2\rangle.
\end{align}
So we have to find the second-order moment of $\partial_k E'$. We can obtain that
\begin{align}
    (\partial_k E')^2&=-\alpha^2\beta^2\bra{0}[R, H]\ket{0}\bra{0}[R, H]\ket{0}+2i\alpha\beta^3\bra{0}[R, H]\ket{0}(\partial_k E)+\beta^4(\partial_k E)^2.
\end{align}
Because of the Equation~\eqref{eq: weingarten}, the Haar measure of the second part is zero. And the third part is the result of Equation~\eqref{eq: bp_variance}. So we only need to calculate the Haar measure about the first part. We use the symbol that $\rho=\ket{\mathbf{init}}\bra{\mathbf{init}}$, the first part will be
\begin{align}
    &\int d\mu\cdot\bra{\mathbf{init}}[R,H]\ket{\mathbf{init}}\bra{\mathbf{init}}[R,H]\ket{\mathbf{init}} \\
    =&\int d\mu\cdot\Tr{\rho(R^\dagger H-HR)\rho(R^\dagger H-HR)} \notag \\
    =&\Tr{\rho\int d\mu\cdot(R^\dagger H\rho R^\dagger H+HR\rho HR)} \notag \\
    -&2\Tr{\rho\int d\mu\cdot HR\rho R^\dagger H}. \notag
\end{align}
Observing this equation, we can notice that the first and second parts are zero, because of Equation~\eqref{eq: weingarten}. So we need to calculate the third Haar measure. By using the conclusion of unitary $1$-design of Equation~\eqref{eq: haar_1}, the third part can be calculated as follows,
\begin{align}
    &\Tr{\rho\int d\mu\cdot HR\rho R^\dagger H}\\
    =&\Tr{\rho\int d\mu_-\int d\mu_+\cdot HU_+V_kU_-\rho (U_+V_kU_-)^\dagger H} \notag \\
    =&\Tr{H\rho H\int d\mu_-\cdot \left[\int d\mu_+\cdot U_+V_kU_-\rho U_-^\dagger V_k^\dagger U_+^\dagger \right]} \notag \\
    =&\Tr{H\rho H\int d\mu_-\cdot \left[\frac{1}{N}\Tr{V_kU_-\rho U_-^\dagger V_k^\dagger}\right]} \notag \\
    =&\Tr{H\rho H\cdot \left[\frac{1}{N}\Tr{\rho}\right]} \notag \\
    =&\frac{1}{N}\Tr{H^2\rho}\Tr{\rho}. \notag
\end{align}

Upon consolidating all the above equations, we can arrive at the final result that
\begin{align}
\mathrm{Var}[\partial_k E']=\frac{2\alpha^2 \beta^2}{N}\Tr{H^2\rho}\Tr{\rho}+\beta^4\langle (\partial_k E)^2 \rangle.
\end{align}

\section{Detail of Eliminate Auxiliary Qubits}
\label{app: eliminate}

In the parameterized layers, we use the same assumptions as before, except that here we have set $\alpha = \beta = \frac{1}{\sqrt{2}}$. Since for most of the training process, the number of pending layers is relatively small compared to the number of adjustable and fixed layers, and since this part contains parameters, we integrate this section into the adjustable layers for computational convenience. $\boldsymbol{\varphi}$ in fixed layers represents a set of fixed parameters exempt from training. The unitary operator of this assumption is $\frac{1}{\sqrt{2}}U'(\boldsymbol{\varphi})(I+U(\boldsymbol{\theta}))$. So, the expectation of $E''$ is 
\begin{align}
    E''(\boldsymbol{\theta}) =& \frac{1}{2}\langle{\mathbf{init}}|(I+U(\boldsymbol{\theta}))^\dagger U'(\varphi)^\dagger H U'(\varphi) (I+U(\boldsymbol{\theta}))|\mathbf{init}\rangle \\
    =&\Tr{\frac{\rho}{2} (I+U(\boldsymbol{\theta}))^\dagger H_\varphi(I+U(\boldsymbol{\theta}))} \notag \\
    =&\Tr{\frac{\rho}{2} (H_\varphi+U^\dagger H_\varphi+H_\varphi U +U^\dagger H_\varphi U)}, \notag
\end{align}
where $H_\varphi=U'(\boldsymbol{\varphi})^\dagger H U'(\boldsymbol{\varphi})$.

Using the Equation~\eqref{eq: partial_u}, the first part of $E''$ is zero, and the gradient of $E''$ is:
\begin{align}
    \partial_k E'' & = \Tr{\frac{i\rho}{2} (R^\dagger H_\varphi-H_\varphi R +R^\dagger H_\varphi U-U^\dagger H_\varphi R} \\
    &= \frac{i}{2}\Tr{\rho [R, H_\varphi]+H_{\varphi,+}[\rho_-,V_k]}, \notag
\end{align}
where $\rho_-=U_-^\dagger \rho U_-$ and $H_{\varphi,+}=U_+^\dagger H_\varphi U_+$.

As the pending layer is located on the right side of the entire parameterized structure, only the unitary $t$-design of $U_+$ needs to be considered.

When $U_+$ is at least a unitary $1$-design, Equation~\eqref{eq: weingarten} leads the first part to zero and Equation~\eqref{eq: haar_1} acts on the second part. So, the first-order moment can be calculated as follows.
\begin{align}
    \langle \partial _kE''\rangle &=\int d\mu \cdot \partial _kE''\\
    &=\frac{i}{2}\int d\mu_-\int d\mu_+\cdot  \Tr{H_{\varphi,+}[\rho_-,V_k]} \notag \\
    &=\frac{i}{2}\int d\mu_-\cdot \Tr{\int d\mu_+\cdot U_+^\dagger H_{\varphi}U_+[\rho_-,V_k]} \notag \\
    &=\frac{i}{2N}\int d\mu_-\cdot \Tr{H_\varphi}\Tr{[\rho_-,V_k]} = 0. \notag  
\end{align}
This result is obtained because the trace of the commutator is zero.

Before calculating the second-order moment, we first solve for $(\partial _k E'')^2$.
\begin{align}
    (\partial _k E'')^2 =&-\frac{1}{4}\langle{\mathbf{init}}|R^\dagger H_\varphi-H_\varphi R|\mathbf{init}\rangle\langle{\mathbf{init}}|R^\dagger H_\varphi-H_\varphi R|\mathbf{init}\rangle \\
    &-\frac{1}{2} \langle{\mathbf{init}}|R^\dagger H_\varphi-H_\varphi R|\mathbf{init}\rangle\langle{\mathbf{init}}|R^\dagger H_\varphi U- U^\dagger H_\varphi R|\mathbf{init}\rangle \notag \\
    &-\frac{1}{4} \langle{\mathbf{init}}|R^\dagger H_\varphi U- U^\dagger H_\varphi R|\mathbf{init}\rangle\langle{\mathbf{init}}|R^\dagger H_\varphi U- U^\dagger H_\varphi R|\mathbf{init}\rangle \notag \\
    =&\frac{1}{2}\Tr{\rho H_\varphi R\rho R^\dagger H_\varphi}-\frac{1}{4}\Tr{H_{\varphi,+}[\rho_-,V_k]H_{\varphi,+}[\rho_-,V_k]}+others. \notag
\end{align}

In this function, the first part is a unitary $1$-design and the second is a unitary $2$-design. The Haar measure of others is zero by the Equation~\eqref{eq: weingarten}. When $U_+$ is at least a unitary $1$-design, Equation~\eqref{eq: haar_1} acts on the first part.
\begin{align}
    &\int d\mu\cdot \frac{1}{2}\Tr{\rho H_\varphi R\rho R^\dagger H_\varphi} \\
    =& \frac{1}{2}\int d\mu_-\cdot\Tr{\int d\mu_+\cdot\rho H_\varphi R\rho R^\dagger H_\varphi} \notag \\
    =&\frac{1}{2}\int d\mu_-\cdot\Tr{\int d\mu_+\cdot\rho H_\varphi(U_+V_kU_-) \rho (U_+V_kU_-)^\dagger H_\varphi} \notag \\
    =&\frac{1}{2N}\int d\mu_-\cdot \Tr{H_\varphi\rho H_\varphi}\Tr{V_kU_-\rho U_-^\dagger V_k^\dagger } \notag \\
    =&\frac{1}{2N}\Tr{\rho H_\varphi^2}\Tr{\rho} \notag 
\end{align}
And when $U_+$ is at least a unitary $2$-design, Equation~\eqref{eq: haar_2} acts on the second part.
\begin{align}
    \int d\mu\cdot &-\frac{1}{4}\Tr{H_{\varphi,+}[\rho_-,V_k]H_{\varphi,+}[\rho_-,V_k]} \\
    =& -\frac{1}{4}\int d\mu_-\cdot\Tr{\int d\mu_+\cdot H_{\varphi,+}[\rho_-,V_k]H_{\varphi,+}[\rho_-,V_k]} \notag \\
    =&-\frac{1}{4}\int d\mu_-\cdot\Tr{\int d\mu_+\cdot U_+^\dagger H_{\varphi}U_+[\rho_-,V_k]U_+^\dagger H_{\varphi}U_+[\rho_-,V_k]} \notag \\
    =&-\frac{1}{4(N^2-1)}\int  d\mu_-\cdot \left[\Tr{H_\varphi }^2\Tr{[\rho_-,V_k]^2 }+\Tr{H_\varphi^2 }\Tr{[\rho_-,V_k]}^2\right]+O(N^{-2}) \notag \\
    =&-\frac{1}{4(N^2-1)}\Tr{H_\varphi }^2\Tr\langle [\rho_-,V_k]^2 \rangle_{U_-}+O(N^{-2}) \notag
\end{align}

Finally, by using the Equation~\eqref{eq: var}, we can get that
\begin{align}
    \mathrm{Var}[\partial_k E'']=
    \begin{cases} 
        \frac{1}{2N} \Tr{\rho H_\varphi^2}\Tr{\rho}+O(N^{-1}), & \text{if } t=1 \\
        \frac{1}{2N} \Tr{\rho H_\varphi^2}\Tr{\rho}-\frac{1}{4(N^2-1)}\Tr{H_\varphi^2}\Tr\langle[\rho_-,V_k]^2\rangle_{U_-}+O(N^{-2}), & \text{if } t=2 
    \end{cases}
\end{align}

We can observe that the result obtained after eliminating the auxiliary qubits is almost identical to that of the entanglement method. This makes it easier to achieve a good gradient as the number of qubits increases. 

\end{document}